\documentclass[aps,pre,12pt]{revtex4-1}
\usepackage{graphicx}
\usepackage{epstopdf}
\usepackage{amsmath}
\usepackage{amssymb}
\usepackage{default}
\usepackage{natbib}
\usepackage{default}
\newcommand{\be}{\begin{equation}}
\newcommand{\bea}{\begin{eqnarray}}
\newcommand{\bc}{\begin{center}}            
\newcommand{\ee}{\end{equation}}
\newcommand{\eea}{\end{eqnarray}}
\newcommand{\ec}{\end{center}}

\newcommand{\baa}{\begin{eqnarray*}}
\newcommand{\eaa}{\end{eqnarray*}}
\begin{document}
\title{Global Linear-irreversible Principle for Optimization
in Finite-time Thermodynamics}
\author{Ramandeep S. Johal}
\email{rsjohal@iisermohali.ac.in}
\affiliation{ Department of Physical Sciences, \\ 
Indian Institute of Science Education and Research Mohali,
Sector 81, S.A.S. Nagar, Manauli PO 140306, Punjab, India}
\begin{abstract}
There is intense effort into understanding the universal properties
of finite-time models of thermal machines---at optimal
performance---such as efficiency at maximum power, coefficient
of performance at maximum cooling power, and other such criteria. 
In this letter, a {\it global} principle consistent
with linear irreversible thermodynamics is proposed  
for the whole cycle---without considering details of irreversibilities 
in the individual steps of the cycle.
This helps to express the total duration of the cycle as 
$\tau \propto {\bar{Q}^2}/{\Delta_{\rm tot} S}$,
where $\bar{Q}$ models the effective heat transferred 
through the machine during the cycle,
 and  $\Delta_{\rm tot} S$ is the total entropy generated.
By taking $\bar{Q}$ in the form of simple algebraic means (such as arithmetic
and geometric means) over the heats exchanged by the reservoirs,
the present approach is able to predict various standard expressions for 
figures of merit at optimal performance, as well as the bounds respected by them.  
It simplifies the optimization procedure to a one-parameter 
optimization, and provides a fresh perspective on the issue 
of universality at optimal performance, for small difference 
in reservoir temperatures. As an illustration, we compare 
performance of a partially optimized four-step endoreversible 
cycle with the present approach.
\end{abstract}
\maketitle
{\it Introduction:}
As the demand of human civilization for useful energy grows,
it becomes more urgent to understand and improve the performance of 
our energy-conversion devices.
Currently, there is a lot of interest 
in characterizing the optimal performance of machines operating in 
finite-time cycles \cite{Curzon1975, Berry1984, Chen1989,
Bejan1996, Salamon2001,Broeck2005, Schmiedl2008, Esposito2010,
Andresen2011, Roco2012b,Roco2012, Wangtu2012pre, Wangtu2012epl,
Wangtu2013epl, Broeck2013, Apertet2013,Correa2014,
Holubec2016, Roco2017,Johal2017}.
As a paradigmatic model, a heat cycle is studied 
between two heat reservoirs at temperatures $T_h$
and $T_c (<T_h)$, whose performance may be optimized 
using a specific objective function, such as:
power output \cite{Curzon1975}, cooling power \cite{Apertet2013}, 
certain trade-off functions between energy gains 
and losses \cite{Angulo1991, Roco2017}, and 
net entropy generation \cite{Bejan1995}.
An important quantity in this regard is 
the figure of merit, such as
efficiency, in case of heat engines, and 
the coefficient of performance (COP) in the 
case of refrigerators. Notably, 
 the bounds on their values predicted 
 by simple, thermodynamic models,  
provide a benchmark for the observed performance 
of real power plants 
\cite{Curzon1975, Bejan1996, Esposito2010, Roco2012b, Johalepjst}.

A major focus has been to understand
whether these figures of merit display universal properties
at optimal performance. For example, the efficiency 
at maximum power (EMP) is often found to be 
equal to, or closely approximated by the elegant expression,  
known as Curzon-Ahlborn (CA) efficiency 
\cite{Reitlinger, Chambadal, Novikov, Curzon1975} 
\be
\eta_{\rm CA} = 1 - \sqrt{1-\eta_{\rm C}},
\ee
where $\eta_{\rm C} = 1-T_c/T_h$ is the Carnot efficiency.
Other expressions for EMP have also been obtained from different models 
\cite{Chen1989, Broeck2005, Schmiedl2008, Tu2008, Esposito2010, Wangtu2012pre, JohalRai2016},
some of them sharing a common universality with CA efficiency, i.e.
for small differences in reservoir temperatures, EMP can be 
written as: $\eta = \eta_{\rm C}/2 + \eta_{\rm C}^2/8 + {\cal O}[\eta_{\rm C}^3]$.
The first-order term arises with strong coupling under linear irreversible 
thermodynamics (LIT) \cite{Broeck2005}, while  
the second-order term is beyond linear response, and has been related to a certain
symmetry property in the model \cite{Lindenberg2009, JohalRai2016}.

The standard analysis often involves solving a two-parameter optimization
problem over, say, the pair of intermediate temperatures of the working 
medium \cite{Curzon1975, Chen1989}, or, the time intervals of contacts 
with reservoirs \cite{Schmiedl2008, Esposito2010}.
In this letter, I formulate a simpler optimization problem for 
finite-time machines. While simplicity might lead to
a certain loss of predictive power, the generality and unifying power of 
the proposed framework are remarkable. Here, rather than applying LIT 
locally, say, at each thermal
contact, I assume a global validity of this principle, i.e. for the 
complete cycle. Accordingly, we need not assume stepwise details of
the cycle, but may simply consider the machine as an
irreversible channel with an effective (thermal) conductivity $\lambda$. 
Quite remarkably, we will recover many of the well-known expressions of the figures
of merit for both engine and refrigerator modes, indicating
a very different origin for them which is    
independent of the physical model, or the processes assumed
in regular models. In this sense, 
the present approach brings together the results from different
models under one general principle consistent with LIT.

Now, according to LIT \cite{Callenbook, Grootbook},
the rate of entropy generation is $\dot{S} = \sum_{\alpha} q_{\alpha}  F_{\alpha}$,  
the sum of products of each flux $q_{\alpha}$ and its associated thermodynamic
force, or affinity $F_{\alpha}$. In the simple case 
of a heat flux $q$ between two heat reservoirs, the corresponding 
thermodynamic force may be defined as: $F = T_{c}^{-1} - T_{h}^{-1}$.
Then, assuming a linear flux-force relation, $q = \lambda F$,
where $\lambda$ is the heat-transfer coefficient, and 
the bilinear form for the rate of entropy generation,
we can write $\dot{S} = q F =  q^2/\lambda$.
Now, if the time interval of heat flow is considered
long enough, then the flux  may be
approximated to be constant over this interval. So the amount of heat
transferred within time $\tau$ is: $Q = q \tau$, which implies  
$\dot{S} = Q^{2} / \lambda \tau^2$. 
Then, the cyclic operation of the machine can be based on
the following two assumptions:

(i) there is an effective heat flux ($\bar{q}$) through the machine over one
  cycle, and the total entropy generation per cycle obeys principles of LIT.

(ii) $\bar{q}$ is determined by an effective, mean value of the heat ($\bar{Q}$) passing 
   from the hot to the cold reservoir in total cycle time $\tau$.
   Therefore, $\bar{q} = \bar{Q}/\tau$.

Assumption (i) implies, $\dot{S}_{\rm tot} = \bar{q}^2 /\lambda$.
Then, the total entropy generation per cycle is: 
$\Delta_{\rm tot} S = \dot{S}_{\rm tot} \tau = \bar{q}^2 \tau /\lambda$.   
From assumption (ii), we have $\Delta_{\rm tot} S = \bar{Q}^2 /\lambda \tau$.
In other words, we can express the total period as:
\be
\tau = \frac{\bar{Q}^2}{\lambda \Delta_{\rm tot} S}.
\label{taudef}
\ee

{\it Optimal power output}: Now, to motivate the optimization procedure, 
we first optimize the power output of a heat engine.
Let $Q_h$ and $Q_c$ be the 
amounts of heat exchanged by the working medium
with the hot ($h$) and cold ($c$) reservoirs. 
Let $W= Q_h - Q_c$ be the total work output 
in a cycle of duration $\tau$. The total entropy generation
per cycle is:
\be
\Delta_{\rm tot} S = -\frac{Q_h}{T_h} +  \frac{Q_c}{T_c} >0.
\label{dst}
\ee
Then the average power output of the cycle
is defined as $P = W/\tau$. Using Eq. (\ref{taudef}), we have
\be
P = \lambda (Q_h -Q_c) \frac{\Delta_{\rm tot} S}{\bar{Q}^2}.
\label{power}
\ee
Introducing the parameters $\nu = Q_c/Q_h$ and $\theta = T_c/T_h$,
and using Eq. (\ref{dst}), we can define a dimensionless power function:
\be 
\tilde{P} \equiv \frac{T_c P}{\lambda} = 
(1-\nu)(\nu-\theta)\frac{Q_{h}^{2}}{\bar{Q}^2}.
\label{tilp}
\ee
For engines, from the positivity of $\Delta_{\rm tot} S$,
we have $\nu > \theta$.
Further, knowledge of a specific value of $\lambda$
is not relevant for performing the optimization. 
However, the above target function is still not in a useful form, until
we specify the form of $\bar{Q}$. 
We assume $\bar{Q}$ 
to be bounded as: $ Q_c \le \bar{Q} \le Q_h$.
Let us analyze these two limits separately. 

${\rm L_h}$) When $\bar{Q} \to Q_h$, 
Eq. (\ref{tilp}) is simplified to: 
$\tilde{P} = (1-\nu)(\nu-\theta)$,
which becomes optimal ($\partial \tilde{P}/ \partial \nu =0 $)
at $\nu  = (1+\theta)/2$. Thus the EMP in this limit is $\eta_l = \eta_{\rm C}/2$.
 This formula is obtained  
when the dissipation at the cold contact is much more important
than at the hot contact \cite{Chen1989, Schmiedl2008, Esposito2010,
Wangtu2012pre, JohalRai2016}. In our model, this limit implies
that when the effective amount of heat passing through the machine 
approaches the heat absorbed from the hot reservoir, then 
the efficiency approaches its lower bound. 
 
${\rm L_c}$) When $\bar{Q} \to Q_c$, then 
$\tilde{P} = (1-\nu)(\nu-\theta){\nu}^{-2}$,
whose optimal value is obtained at $\nu = 2\theta/(1+\theta)$, 
or the EMP in this case is $ \eta_u = \eta_{\rm C}/(2-\eta_{\rm C})$.
Again, this formula is obtained 
as the upper bound for efficiency  \cite{Chen1989, Schmiedl2008,
Esposito2010, Wangtu2012pre, JohalRai2016}, when the dissipation
at the cold contact approaches the reversible limit, or,  
is negligible in comparison to dissipation at the hot contact.
In our model, $\bar{Q} \to Q_c$ implies that as the effective
heat through the machine reaches its lowest possible value $Q_c$,
the efficiency approaches its upper bound. 

Now, it seems natural to assume that, in general,
$\bar{Q} \equiv \bar{Q}(Q_h,Q_c)$ may be taken as 
a mean value \cite{Bullen2003}, interpolating
between $Q_c$ and $Q_h$. In the following, 
we explore consequences of making some simple choices
of these mean values. It will be seen that 
the mean being a homogeneous function 
of the first degree in its arguments, implies 
the condition $\bar{Q}(Q_h,Q_c) = Q_h \bar{Q}(1,\nu)$,
and so the maximization of the power output is reduced
to a simple one-parameter optimization problem.

Let $\bar{Q}$ be given by a weighted 
arithmetic mean: $\bar{Q} = \omega Q_h + (1-\omega) Q_c$,
where the weight  $0\le \omega \le 1$.
With this form, Eq. (\ref{tilp}) is explictly given by:
\be
\tilde{P}(\nu) = \frac{(1-\nu)(\nu-\theta)}{[\omega +(1-\omega)\nu]^2}.
\label{preduce}
\ee
Then the optimum of $\tilde{P}$ is  obtained at:
$\tilde{\nu} = [{2 \theta + \omega (1-\theta)}]/[{1+\theta + \omega (1-\theta)}]$.
The maximal nature of the optimum can be ascertained: 
$\left. \partial^2 \tilde{P}/ \partial \nu^2 \right|_{\nu = \tilde{\nu}} <0$.
As a result, EMP, $\tilde{\eta} = 1-\tilde{\nu}$, is found to be:
\be
\tilde{\eta} = \frac{\eta_{\rm C}}{2-(1-\omega)\eta_{\rm C}}.
\label{empw}
\ee
The above form has been obtained in Refs. 
\cite{Chen1989, Schmiedl2008, Esposito2010, Wangtu2012pre, JohalRai2016},
where the parameter $\omega$ may be determinable, 
for example, in terms of the ratio of the dissipation constants
or thermal conductivities of the thermal contacts \cite{Chen1989, Esposito2010}.
In the present approach, the parameter 
$0\le \omega \le 1$ is undetermined. 
In the absence of additional information, one may
choose equal weights ($\omega =1/2$), or the 
{\it symmetric} arithmetic mean $\bar{Q} = (Q_h+ Q_c)/2$.
This results in the so-called Schmiedl-Seifert (SS)
efficiency $\tilde{\eta}_{\rm SS} = {2\eta_{\rm C}}/(4-\eta_{\rm C})$ 
\cite{Orlov, Schmiedl2008}.

\begin{figure}[ht]
  \centering
    \includegraphics[width=10cm,height=10cm,keepaspectratio]{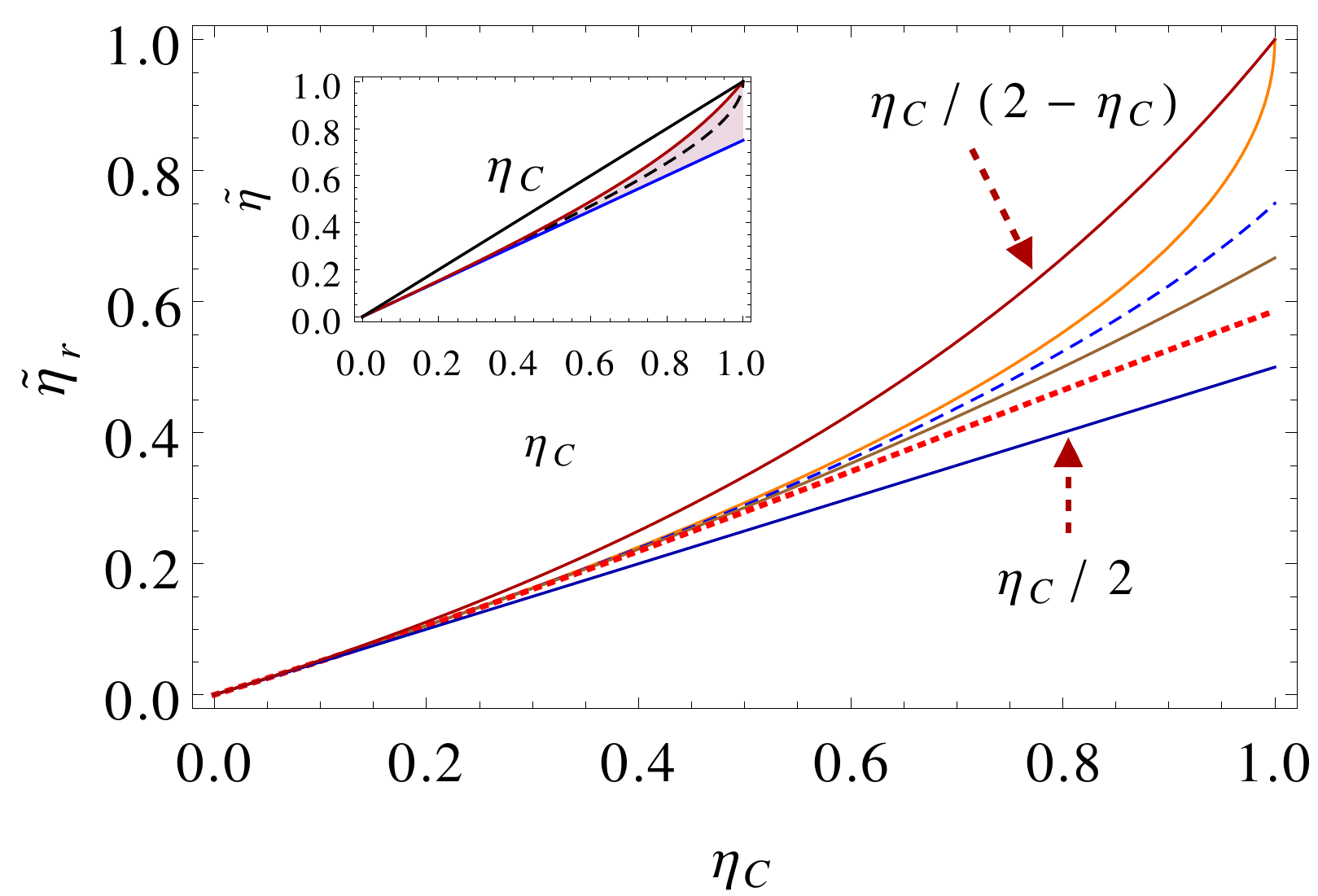}
    \caption{EMP, $\tilde{\eta}_r$,  versus Carnot efficiency 
    $\eta_{\rm C}$, for different values of the parameter $r$ 
    of the generalized mean. The upper (lower) bound
    is obtained with $r$ approaching $-\infty (+\infty)$. 
    In between these bounds, from top to bottom, $r= 0, 1/2, 1, 2$
    respectively, depicting that for a given ratio of 
    reservoir temperatures, $\theta$, EMP decreases
    monotonically with $r$. Inset: The shaded region
    depicts the efficiency at optimal ecological criterion,  
     bounded from below and above as in Eq. (\ref{etaeco}).
    The dashed curve inside the shaded region, depicts Eq. (\ref{etaangulo}),
     obtained with the geometric-mean value $\overline{Q} = \sqrt{Q_h Q_c}$.
    }
    \label{fig1}
  \end{figure}

As an alternative choice, if we set $\bar{Q} = \sqrt{Q_h Q_c}$, i.e. 
the geometric mean of $Q_h$ and $Q_c$, then we obtain 
$\tilde{P} = (1-\nu)(\nu-\theta)/\nu$, which becomes optimal 
at CA efficiency, $\tilde{\eta} =\eta_{\rm CA}$. 
Further,  we may use a generalized, symmetric mean defined as: 
$\bar{Q} = [(Q_{h}^{r} + Q_{c}^{r})/2]^{1/r}$,
where $r$ is a real parameter \cite{Hardy1952, Bullen2003}. 
Special cases with $r=-1,0,1,2$ corrrespond
respectively to harmonic, geometric, arithmetic, and quadratic means.
The dimensionless power output is then given by:
\be
\tilde{P}(\nu) = \frac{(1-\nu)(\nu-\theta)}{[(1 +\nu^r)/2]^{2/r}},
\ee
whose optimum is determined by following condition:
\be
(1+\theta){\tilde{\nu}}^r -2 \theta {\tilde{\nu}}^{r-1} 
+ 2 \tilde{\nu} - (1+\theta) = 0.
\label{nur}
\ee
The above equation cannot be analytically solved for $\tilde{\nu} =  
1 - \tilde{\eta}_r$, with general $r$.
However,  it can be shown from (\ref{nur})
that $\partial \tilde{\nu}/\partial r > 0$, which implies
that EMP $\tilde{\eta}_r$ decreases monotonically with increasing $r$.
In particular, when $r\to  +\infty (-\infty)$, then $\bar{Q} \to Q_h (Q_c)$
and so we have $\tilde{\eta}_{\infty} \to \eta_{\rm C}/2$, and   
$\tilde{\eta}_{-\infty} \to \eta_{\rm C}/(2-\eta_{\rm C})$, the lower and upper
bounds of EMP discussed earlier. Incidentally, 
this helps to notice that CA efficiency ($r \to 0$) is
 higher than SS efficiency ($r=1$), for a given $\theta$ (see Fig. \ref{fig1}). 

For small numerical values of $r$, we can look into the general
form of efficiency by assuming $\tilde{\eta}_r = a_1 \eta_{\rm C} + a_2 \eta_{\rm C}^{2} 
+ a_3 \eta_{\rm C}^{3}$, close to equilibrium, where we have $\eta_{\rm C}$ as the small parameter.
Substituting this form in Eq. (\ref{nur}), and keeping terms upto ${\cal O}( \eta_{\rm C}^{3})$,
we determine the coefficients $a_i$. As a result, for small temperature differences,
we have 
\be 
\tilde{\eta}_r = \frac{\eta_{\rm C}^{}}{2} + \frac{\eta_{\rm C}^{2}}{8} 
+ \frac{2-r}{32}\eta_{\rm C}^{3}.
\label{erseries}
\ee
Thus, we observe that with some well-known symmetric means, the first two 
terms in the expansion of EMP have the same universal coefficients
as were earlier obtained within LIT \cite{Broeck2005, Lindenberg2009}. 
Interestingly, the third-order term in Eq. (\ref{erseries}) also matches 
with a similar expansion of EMP recently found for Carnot engines
within a perturbative approach for open quantum systems  \cite{Cavina2017}.
Thereby, the parameter $r$ is analogous to the 
frequency scaling exponent of the spectral density of the heat reservoirs.

So, assuming the global validity of LIT for a heat cycle, 
along with a specific form of the effective heat transferred between
the two heat reservoirs, 
leads to a one-parameter optimization problem, whereby 
the EMP matches with the well-known expressions predicted 
by more elaborate models. This is the main result of the present letter. 
In the following, we extend the analysis to other target functions, 
and derive various known forms of the figures of merit. 
{\it Optimal ecological criterion}: 
Although, maximizing power output may be regarded as
a reasonable goal for a finite-time engine, we note
that such an engine also leads to irreversible entropy
generation. So, for a cleaner production of useful energy,
the so-called ecological criterion provides  
a good compromise between energy benefits and losses in thermal machines 
\cite{Angulo1991, Yan1993comment}. For the specific case of 
a two heat-reservoir setup, it is also equivalent to 
the so-called $\dot{\Omega}$-criterion \cite{Calvo2001, Roco2017}.  
For a heat engine, the criterion is given by: 
$\dot{\Omega} = (2 \eta - \eta_{\rm C}){Q_h}/{\tau}$, 
or, in dimensionless form 
\be
\tilde{\dot{\Omega}} = \frac{T_c \dot{\Omega}}{\lambda} =
\frac{(1+\theta -2 \nu)(\nu-\theta)}{[\omega +(1-\omega)\nu]^2}.
\ee
Again, to obtain the optimal working condition, 
we set $\partial {\tilde{\dot{\Omega}}}/ \partial \nu =0$, and thus 
obtain the efficiency at optimal ${\dot{\Omega}}$ as \cite{Roco2017}:
\be 
\tilde{\eta} = \frac{3-2\eta_{\rm C} +2 \omega \eta_{\rm C}}
{4- 3\eta_{\rm C} + 3\omega \eta_{\rm C}}\eta_{\rm C}.
\label{ecoeta}
\ee
The upper and lower bounds of (\ref{ecoeta}) are given by
\be
\frac{3}{4} \eta_{\rm C} \le  \tilde{\eta} \le  
\frac{3-2 \eta_{\rm C}}{4- 3\eta_{\rm C}}\eta_{\rm C} = \eta^*.
\label{etaeco}
\ee
For small temperature differences, the upper bound $\eta^*$
behaves as ${\eta}^* = {3} \eta_{\rm C}/4 + {\eta_{\rm C}^{2}}/{16}
 + {\cal O}(\eta_{\rm C}^{3})$.

On the other hand, 
the use of geometric mean for the case of ecological criterion for an engine 
yields
$\tilde{\dot{\Omega}} = (1+\theta - 2 \nu)(\nu-\theta)/{\nu}$, whose
optimum is obtained at 
$\nu = \sqrt{\theta(1+\theta)/2}$. So the efficiency at optimal 
ecological criterion is then given by \cite{Yan1993comment, Long2015}:
\be
\tilde{\eta} = 1-\sqrt{\frac{(1-\eta_{\rm C})(2-\eta_{\rm C})}{2}}.
\label{etaangulo}
\ee
In this case, for small temperature differences, the leading order terms are 
given as $\tilde{\eta} = {3}\eta_{\rm C}/4 + {\eta_{\rm C}^{2}}/{32}
 + {\cal O}(\eta_{\rm C}^{3})$. The exact behavior is plotted as the dashed
 line within the inset, in Fig. 1.
 
Next, we consider the operation of a heat cycle as a refrigerator.
For refrigerators, the coefficient of performance (COP) is given by 
$\xi = {\cal Q}_c/{\cal W} = \nu/(1-\nu)$, where $\nu = {\cal Q}_c/{\cal Q}_h$,
and the Carnot coefficient is $\xi_{\rm C} = \theta/(1-\theta)$.
Here, ${\cal Q}_c$ is the heat extracted from
the cold reservoir and driven into the hot reservoir by an input
of ${\cal W}$ amount of work, in a cycle of time period $\tau$.
The total entropy generation per cycle is: 
\be
\Delta_{\rm tot} {\cal S} = \frac{ {\cal Q}_h}{T_h} 
-  \frac{ {\cal Q}_c}{T_c} >0.
\label{dstr}
\ee
Now, for irreversible refrigerators, the choice of optimization
criteria that may be analogous to heat engines, is not straightforward. 
For instance, the cooling power or the rate of refrigeration, 
${\cal Q}_c /\tau$,  may not 
have an optimum for certain models \cite{Chen1990, Apertet2013}. 
Below, we illustrate the optimization with some of the proposed choices.

{\it Optimal $\chi$-criterion}:
The so-called $\chi$-criterion is defined as \cite{Chen1990, Roco2012}  
$\chi = {\xi Q_c}/{\tau}$. This criterion simultaneously considers 
the COP and the cooling power, due to 
a certain complementarity between the two quantities, i.e. if we maximize
one, it minimizes the other. Defining, $\tilde{\chi} = T_c \chi/\lambda$,
we have:
\be
\tilde{\chi} = \frac{\nu^2(\theta-\nu)}{(1-\nu)[\omega +(1-\omega)\nu]^2}.
\label{tilchi}
\ee
Note that for an irreversible refrigerator, we have $\theta > \nu$.
Then,
setting $\partial \tilde{\chi}/ \partial \nu =0$, 
we can obtain the COP at optimal performance:
\be
\tilde{\xi} = \frac{\sqrt{\omega}}{2} 
           \left( \sqrt{9 \omega + 8 \xi_{\rm C}} - 3 \sqrt{\omega}\right).
\ee
As $0 \le \omega \le 1$, it implies the bounds on COP as:
\be
0 \le \tilde{\xi} \le \frac{1}{2} 
           \left( \sqrt{9 + 8 \xi_{\rm C}} - 3 \right),
           \ee
which have been earlier obtained in the literature, from a two-parameter
optimization procedure. Note that, here we obtain a simple
closed form for $\tilde{\xi}$, which has not been possible
in the low-dissipation approach of Ref. \cite{Roco2012}.

{\it Optimal cooling power}:
The target function for a refrigerator may be chosen as
the cooling power $Z = Q_c/\tau$, or in dimensionless form 
\be 
\tilde{Z} = \frac{T_c Z}{\lambda} = 
\frac{\nu (\theta-\nu)}{[\omega +(1-\omega)\nu]^2}.
\ee
Then, setting $\partial \tilde{Z}/ \partial \nu =0$, 
we obtain COP at optimal $Z$ to be:
\be
\tilde{\xi} = \frac{\omega \xi_{\rm C}}{2 \omega + \xi_{\rm C}}.
\label{copcool}
\ee
The above expression is exactly the one derived in Ref. \cite{Apertet2013}
for the so-called exoreversible refrigerator,
with the consequent bounds on COP as:  $0 \le \tilde{\xi} 
\le \xi_{\rm C}/(2+ \xi_{\rm C})$.

When the ecological or the trade-off criterion for the refrigerator \cite{Calvo2001}
is implemented as: ${\dot{\Omega}}_R = (2\xi - \xi_{\rm C}) {\cal W}/\tau$, 
 with $\bar{Q}$ as a geometric mean, then 
we obtain the COP at optimal $\dot{{\Omega}}_R$ 
\cite{Calvo2001,Calvo2013}, as: 
\be
\tilde{\xi} = \frac{\xi_{\rm C}}{\sqrt{(1+\xi_{\rm C})(2+\xi_{\rm C})}-\xi_{\rm C}}.
\label{xicr}
\ee
Similarly, it can be seen that for a refrigerator and with geometric mean, we obtain 
$\tilde{\chi} = \nu(\theta-\nu)/(1-\nu)$, whose optimal value is obtained at 
$\tilde{\xi} = \sqrt{1+ \xi_{\rm C}} -1$ \cite{Roco2012b}. On the other hand,
the cooling power does not have an optimum if the geometric mean is used.

{\it An illustration:} 
In the following, we compare the above effective approach with the optimization
of power output in a four-step cycle within the so-called endoreversible approximation 
\cite{Curzon1975, Chen1989, Andresen2011}.
The latter implies that the only sources of irreversibilities during the cycle 
happen to be the thermal contacts with the heat reservoirs, when 
heat is exchanged between reservoirs and the working medium across a finite heat conductance.
Consequently, there are two such thermal steps, occuring with time 
intervals $t_h$ and $t_c$. Further, the working medium is assumed
to maintain a fixed temperature during a specific thermal contact,
which we assume to be $T_1$ and $T_2$
   during hot and cold contact respectively ($T_h > T_1 > T_2 > T_c$). 
The other two steps of the cycle are adiabatic in
nature---preserving the entropy of the working medium---and
are assumed to occur with a negligible time interval.

   Now, the problem of power optimization for the above model involves 
   {\it two} variables, which may 
   be conveniently chosen as the two temperatures of the working
   medium. Equivalently, we may choose another pair of variables
   as discussed below. 
   Consider the heat flux between the working medium and the respective 
   reservoir to be given by \cite{Chen1989} 
\bea 
q_h & = & \alpha_h \left ( T_{1}^{-1} - T_{h}^{-1}      \right ), \label{rateh}\\
q_c & = & \alpha_c \left ( T_{c}^{-1} - T_{2}^{-1}      \right ), \label{ratec}
   \eea
where $\alpha_j >0$, with $j=c,h$ are the heat-transfer coefficients.
As the flux is assumed to be constant during the thermal contact, 
so the amounts of heat transferred during the times 
$t_h$ and $t_c$, respectively are:
$Q_h = q_h t_h$ and $Q_c = q_c t_c$.
%
Further, the cyclicity within the working medium implies
${Q_h}/{T_1} = {Q_c}/{T_2}$,
also known as the endoreversibility condition.

The extracted work per cycle is $W = Q_h-Q_c$,
with thermal efficiency equal to
\be
\eta = \frac{W}{Q_h} = 1-\frac{T_2}{T_1} \equiv 1- \nu.
\label{effl}
\ee 
Then, the average power per cycle is defined as 
\be
P = \frac{Q_h - Q_c}{t_h + t_c}.
\ee
which can be expressed as a function of $T_1$ and $T_2$,
or equivalently (from Eq. (\ref{effl})), as a function of 
$T_1$ and $\nu$: $P\equiv P(T_1, \nu)$. 
Rather than optimizing the power w.r.t the two 
variables, let us consider a partial optimization,  
by setting ($\partial P /\partial T_1)_{\nu} =0$, which 
yields the optimum value of $T_1$:
\be
\tilde{T}_1 = \frac{K + {\nu}^{-1}}{K T_{h}^{-1} + T_{c}^{-1}},
\ee
where $K = \sqrt{\alpha_h / \alpha_c}$. Using the above value, we obtain 
${P} (\nu) = P(\tilde{T}_1, \nu)$, as 
\be
{P} (\nu) = \frac{\alpha_h}{T_{c}^{}} \frac{(1-\nu) (\nu-\theta) }{[1+ K \nu ]^2},
\ee
which can be rewritten as:
\be
{P} (\nu) = \frac{\alpha_h \alpha_c}{T_{c}^{} (\sqrt{\alpha_h} + \sqrt{\alpha_c})^2 }
\frac{(1-\nu) (\nu-\theta) }{[\omega + (1-\omega) \nu ]^2},
\ee 
where $\omega = (1+K)^{-1}$. We identify  
 $\lambda = \alpha_h \alpha_c {(\sqrt{\alpha_h} + \sqrt{\alpha_c})^{-2}}$,
and so we can compare $T_c P(\nu)/\lambda$ in the above 
with the reduced power in Eq. (\ref{preduce}).
Thus we note that the partially optimized power output in the endoreversible model 
with the inverse-temperature law, is equivalent to our effective model that employs  
a weighted arithmetic mean for the effective transferred heat. 
Similarly, it can be shown that within the endoreversible model under the 
assumption of Newtonian heat transfer \cite{Curzon1975, Chen1989}, 
\bea 
q_h & = & \alpha_{h}^{'} \left ( T_{h}^{} - T_{1}^{}      \right ), \label{ratehn}\\
q_c & = & \alpha_{c}^{'} \left ( T_{2}^{} - T_{c}^{}      \right ), \label{ratecn}
   \eea
where $\alpha_{j}^{'} >0$ are thermal conductivities,
the partially optimized power is obtained in a form that  
implies the effective heat as a geometric mean $\bar{Q} = \sqrt{Q_h Q_c}$,
with the effective value of  $\lambda = 
T_h T_c \alpha_{h}^{'} \alpha_{c}^{'} {(\sqrt{\alpha_{h}^{'}} + \sqrt{\alpha_{c}^{'}})^{-2}}$.

{\it Concluding remarks and outlook:}  
An effective framework has been proposed for a class of thermal machines 
based on LIT, that makes a novel use of algebraic means 
\cite[and references therein]{Johal2016} to model
the effective heat transferred in a cycle.
Quite surprisingly, the method reproduces well-known bounds 
and expressions for figures of merit---both for the engines as well as refrigerators,
 without incorporating details of a specific heat cycle. 
This has been illustrated for various optimization criteria.  
These general results arising from a simple framework, 
thereby suggest  a universal character
of these figures of merit.  

The approach also provides a fresh perspective on the issue 
of universality of EMP near equilibrium 
\cite{Broeck2005, Lindenberg2009, JohalRai2016}, as in Eq. (\ref{erseries}).
The first order term ($\eta_C/2$) is universal, 
for any mean $\bar{Q}$, with extremal values of $Q_h$ and $Q_c$. 
Then, the universality of the second-order term   
can be related to the property of the mean $\bar{Q}$ being symmetric.
Further, an agreement upto third-order term has been seen 
within an open quantum systems framework. 
Thus, the present approach could be relevant for the optimal regimes of 
quantum heat engines, where such efficiencies and bounds 
have been recently derived \cite{Cavina2017}.

It is desirable that the simple approach proposed 
above can be generalized to
include more realistic scenarios, such as
allowing for heat leaks between the reservoirs,
finite sizes of heat source/sink, and the nonlinear regime.
In particular, it is important to understand the physical
reason as to the use of different means leading to varied
expressions for the figures of merit. Here, the comparison
with the partially optimized endoreversible cycle can 
provide a useful insight. 
Finally, it will be good if the proposed approach encourages 
a more global perspective while regarding the 
operation of machines and their impact on our environment. 


\begin{thebibliography}{46}%
\makeatletter
\providecommand \@ifxundefined [1]{%
 \@ifx{#1\undefined}
}%
\providecommand \@ifnum [1]{%
 \ifnum #1\expandafter \@firstoftwo
 \else \expandafter \@secondoftwo
 \fi
}%
\providecommand \@ifx [1]{%
 \ifx #1\expandafter \@firstoftwo
 \else \expandafter \@secondoftwo
 \fi
}%
\providecommand \natexlab [1]{#1}%
\providecommand \enquote  [1]{``#1''}%
\providecommand \bibnamefont  [1]{#1}%
\providecommand \bibfnamefont [1]{#1}%
\providecommand \citenamefont [1]{#1}%
\providecommand \href@noop [0]{\@secondoftwo}%
\providecommand \href [0]{\begingroup \@sanitize@url \@href}%
\providecommand \@href[1]{\@@startlink{#1}\@@href}%
\providecommand \@@href[1]{\endgroup#1\@@endlink}%
\providecommand \@sanitize@url [0]{\catcode `\\12\catcode `\$12\catcode
  `\&12\catcode `\#12\catcode `\^12\catcode `\_12\catcode `\%12\relax}%
\providecommand \@@startlink[1]{}%
\providecommand \@@endlink[0]{}%
\providecommand \url  [0]{\begingroup\@sanitize@url \@url }%
\providecommand \@url [1]{\endgroup\@href {#1}{\urlprefix }}%
\providecommand \urlprefix  [0]{URL }%
\providecommand \Eprint [0]{\href }%
\providecommand \doibase [0]{http://dx.doi.org/}%
\providecommand \selectlanguage [0]{\@gobble}%
\providecommand \bibinfo  [0]{\@secondoftwo}%
\providecommand \bibfield  [0]{\@secondoftwo}%
\providecommand \translation [1]{[#1]}%
\providecommand \BibitemOpen [0]{}%
\providecommand \bibitemStop [0]{}%
\providecommand \bibitemNoStop [0]{.\EOS\space}%
\providecommand \EOS [0]{\spacefactor3000\relax}%
\providecommand \BibitemShut  [1]{\csname bibitem#1\endcsname}%
\let\auto@bib@innerbib\@empty
\bibitem [{\citenamefont {Curzon}\ and\ \citenamefont
  {Ahlborn}(1975)}]{Curzon1975}%
  \BibitemOpen
  \bibfield  {author} {\bibinfo {author} {\bibfnamefont {F.~L.}\ \bibnamefont
  {Curzon}}\ and\ \bibinfo {author} {\bibfnamefont {B.}~\bibnamefont
  {Ahlborn}},\ }\href@noop {} {\bibfield  {journal} {\bibinfo  {journal} {Am.
  J. Phys.}\ }\textbf {\bibinfo {volume} {43}},\ \bibinfo {pages} {22}
  (\bibinfo {year} {1975})}\BibitemShut {NoStop}%
\bibitem [{\citenamefont {Andresen}\ \emph {et~al.}(1984)\citenamefont
  {Andresen}, \citenamefont {Salamon},\ and\ \citenamefont
  {Berry}}]{Berry1984}%
  \BibitemOpen
  \bibfield  {author} {\bibinfo {author} {\bibfnamefont {B.}~\bibnamefont
  {Andresen}}, \bibinfo {author} {\bibfnamefont {P.}~\bibnamefont {Salamon}}, \
  and\ \bibinfo {author} {\bibfnamefont {R.~S.}\ \bibnamefont {Berry}},\
  }\href@noop {} {\bibfield  {journal} {\bibinfo  {journal} {Phys. Today}\
  }\textbf {\bibinfo {volume} {37}},\ \bibinfo {pages} {62} (\bibinfo {year}
  {1984})}\BibitemShut {NoStop}%
\bibitem [{\citenamefont {Chen}\ and\ \citenamefont {Yan}(1989)}]{Chen1989}%
  \BibitemOpen
  \bibfield  {author} {\bibinfo {author} {\bibfnamefont {L.}~\bibnamefont
  {Chen}}\ and\ \bibinfo {author} {\bibfnamefont {Z.}~\bibnamefont {Yan}},\
  }\href@noop {} {\bibfield  {journal} {\bibinfo  {journal} {J. Chem. Phys.}\
  }\textbf {\bibinfo {volume} {90}},\ \bibinfo {pages} {3740} (\bibinfo {year}
  {1989})}\BibitemShut {NoStop}%
\bibitem [{\citenamefont {Bejan}(1996)}]{Bejan1996}%
  \BibitemOpen
  \bibfield  {author} {\bibinfo {author} {\bibfnamefont {A.}~\bibnamefont
  {Bejan}},\ }\href@noop {} {\bibfield  {journal} {\bibinfo  {journal} {J.
  Appl. Phys.}\ }\textbf {\bibinfo {volume} {79}},\ \bibinfo {pages} {1191}
  (\bibinfo {year} {1996})}\BibitemShut {NoStop}%
\bibitem [{\citenamefont {Salamon}\ \emph {et~al.}(2001)\citenamefont
  {Salamon}, \citenamefont {Nulton}, \citenamefont {Siragusa}, \citenamefont
  {Andersen},\ and\ \citenamefont {Limon}}]{Salamon2001}%
  \BibitemOpen
  \bibfield  {author} {\bibinfo {author} {\bibfnamefont {P.}~\bibnamefont
  {Salamon}}, \bibinfo {author} {\bibfnamefont {J.}~\bibnamefont {Nulton}},
  \bibinfo {author} {\bibfnamefont {G.}~\bibnamefont {Siragusa}}, \bibinfo
  {author} {\bibfnamefont {T.}~\bibnamefont {Andersen}}, \ and\ \bibinfo
  {author} {\bibfnamefont {A.}~\bibnamefont {Limon}},\ }\href
  {http://EconPapers.repec.org/RePEc:eee:energy:v:26:y:2001:i:3:p:307-319}
  {\bibfield  {journal} {\bibinfo  {journal} {Energy}\ }\textbf {\bibinfo
  {volume} {26}},\ \bibinfo {pages} {307} (\bibinfo {year} {2001})}\BibitemShut
  {NoStop}%
\bibitem [{\citenamefont {Van~den Broeck}(2005)}]{Broeck2005}%
  \BibitemOpen
  \bibfield  {author} {\bibinfo {author} {\bibfnamefont {C.}~\bibnamefont
  {Van~den Broeck}},\ }\href {\doibase 10.1103/PhysRevLett.95.190602}
  {\bibfield  {journal} {\bibinfo  {journal} {Phys. Rev. Lett.}\ }\textbf
  {\bibinfo {volume} {95}},\ \bibinfo {pages} {190602} (\bibinfo {year}
  {2005})}\BibitemShut {NoStop}%
\bibitem [{\citenamefont {Schmiedl}\ and\ \citenamefont
  {Seifert}(2008)}]{Schmiedl2008}%
  \BibitemOpen
  \bibfield  {author} {\bibinfo {author} {\bibfnamefont {T.}~\bibnamefont
  {Schmiedl}}\ and\ \bibinfo {author} {\bibfnamefont {U.}~\bibnamefont
  {Seifert}},\ }\href@noop {} {\bibfield  {journal} {\bibinfo  {journal}
  {Europhys. Lett.}\ }\textbf {\bibinfo {volume} {81}},\ \bibinfo {pages}
  {20003} (\bibinfo {year} {2008})}\BibitemShut {NoStop}%
\bibitem [{\citenamefont {Esposito}\ \emph {et~al.}(2010)\citenamefont
  {Esposito}, \citenamefont {Kawai}, \citenamefont {Lindenberg},\ and\
  \citenamefont {Van~den Broeck}}]{Esposito2010}%
  \BibitemOpen
  \bibfield  {author} {\bibinfo {author} {\bibfnamefont {M.}~\bibnamefont
  {Esposito}}, \bibinfo {author} {\bibfnamefont {R.}~\bibnamefont {Kawai}},
  \bibinfo {author} {\bibfnamefont {K.}~\bibnamefont {Lindenberg}}, \ and\
  \bibinfo {author} {\bibfnamefont {C.}~\bibnamefont {Van~den Broeck}},\ }\href
  {\doibase 10.1103/PhysRevLett.105.150603} {\bibfield  {journal} {\bibinfo
  {journal} {Phys. Rev. Lett.}\ }\textbf {\bibinfo {volume} {105}},\ \bibinfo
  {pages} {150603} (\bibinfo {year} {2010})}\BibitemShut {NoStop}%
\bibitem [{\citenamefont {Andresen}(2011)}]{Andresen2011}%
  \BibitemOpen
  \bibfield  {author} {\bibinfo {author} {\bibfnamefont {B.}~\bibnamefont
  {Andresen}},\ }\href {\doibase 10.1002/anie.201001411} {\bibfield  {journal}
  {\bibinfo  {journal} {Angewandte Chemie International Edition}\ }\textbf
  {\bibinfo {volume} {50}},\ \bibinfo {pages} {2690} (\bibinfo {year}
  {2011})}\BibitemShut {NoStop}%
\bibitem [{\citenamefont {de~Tom\'as}\ \emph {et~al.}(2012)\citenamefont
  {de~Tom\'as}, \citenamefont {Hern\'andez},\ and\ \citenamefont
  {Roco}}]{Roco2012b}%
  \BibitemOpen
  \bibfield  {author} {\bibinfo {author} {\bibfnamefont {C.}~\bibnamefont
  {de~Tom\'as}}, \bibinfo {author} {\bibfnamefont {A.~C.}\ \bibnamefont
  {Hern\'andez}}, \ and\ \bibinfo {author} {\bibfnamefont {J.~M.~M.}\
  \bibnamefont {Roco}},\ }\href {\doibase 10.1103/PhysRevE.85.010104}
  {\bibfield  {journal} {\bibinfo  {journal} {Phys. Rev. E}\ }\textbf {\bibinfo
  {volume} {85}},\ \bibinfo {pages} {010104(R)} (\bibinfo {year}
  {2012})}\BibitemShut {NoStop}%
\bibitem [{\citenamefont {Wang}\ \emph {et~al.}(2012)\citenamefont {Wang},
  \citenamefont {Li}, \citenamefont {Tu}, \citenamefont {Hern\'andez},\ and\
  \citenamefont {Roco}}]{Roco2012}%
  \BibitemOpen
  \bibfield  {author} {\bibinfo {author} {\bibfnamefont {Y.}~\bibnamefont
  {Wang}}, \bibinfo {author} {\bibfnamefont {M.}~\bibnamefont {Li}}, \bibinfo
  {author} {\bibfnamefont {Z.~C.}\ \bibnamefont {Tu}}, \bibinfo {author}
  {\bibfnamefont {A.~C.}\ \bibnamefont {Hern\'andez}}, \ and\ \bibinfo {author}
  {\bibfnamefont {J.~M.~M.}\ \bibnamefont {Roco}},\ }\href {\doibase
  10.1103/PhysRevE.86.011127} {\bibfield  {journal} {\bibinfo  {journal} {Phys.
  Rev. E}\ }\textbf {\bibinfo {volume} {86}},\ \bibinfo {pages} {011127}
  (\bibinfo {year} {2012})}\BibitemShut {NoStop}%
\bibitem [{\citenamefont {Wang}\ and\ \citenamefont
  {Tu}(2012{\natexlab{a}})}]{Wangtu2012pre}%
  \BibitemOpen
  \bibfield  {author} {\bibinfo {author} {\bibfnamefont {Y.}~\bibnamefont
  {Wang}}\ and\ \bibinfo {author} {\bibfnamefont {Z.~C.}\ \bibnamefont {Tu}},\
  }\href {\doibase 10.1103/PhysRevE.85.011127} {\bibfield  {journal} {\bibinfo
  {journal} {Phys. Rev. E}\ }\textbf {\bibinfo {volume} {85}},\ \bibinfo
  {pages} {011127} (\bibinfo {year} {2012}{\natexlab{a}})}\BibitemShut
  {NoStop}%
\bibitem [{\citenamefont {Wang}\ and\ \citenamefont
  {Tu}(2012{\natexlab{b}})}]{Wangtu2012epl}%
  \BibitemOpen
  \bibfield  {author} {\bibinfo {author} {\bibfnamefont {Y.}~\bibnamefont
  {Wang}}\ and\ \bibinfo {author} {\bibfnamefont {Z.~C.}\ \bibnamefont {Tu}},\
  }\href {http://stacks.iop.org/0295-5075/98/i=4/a=40001} {\bibfield  {journal}
  {\bibinfo  {journal} {EPL (Europhysics Letters)}\ }\textbf {\bibinfo {volume}
  {98}},\ \bibinfo {pages} {40001} (\bibinfo {year}
  {2012}{\natexlab{b}})}\BibitemShut {NoStop}%
\bibitem [{\citenamefont {Yang}\ and\ \citenamefont
  {Zhan-Chun}(2013)}]{Wangtu2013epl}%
  \BibitemOpen
  \bibfield  {author} {\bibinfo {author} {\bibfnamefont {W.}~\bibnamefont
  {Yang}}\ and\ \bibinfo {author} {\bibfnamefont {T.}~\bibnamefont
  {Zhan-Chun}},\ }\href {http://stacks.iop.org/0253-6102/59/i=2/a=08}
  {\bibfield  {journal} {\bibinfo  {journal} {Communications in Theoretical
  Physics}\ }\textbf {\bibinfo {volume} {59}},\ \bibinfo {pages} {175}
  (\bibinfo {year} {2013})}\BibitemShut {NoStop}%
\bibitem [{\citenamefont {{C. Van den Broeck}}(2013)}]{Broeck2013}%
  \BibitemOpen
  \bibfield  {author} {\bibinfo {author} {\bibnamefont {{C. Van den Broeck}}},\
  }\href {\doibase 10.1209/0295-5075/101/10006} {\bibfield  {journal} {\bibinfo
   {journal} {EPL}\ }\textbf {\bibinfo {volume} {101}},\ \bibinfo {pages}
  {10006} (\bibinfo {year} {2013})}\BibitemShut {NoStop}%
\bibitem [{\citenamefont {Apertet}\ \emph {et~al.}(2013)\citenamefont
  {Apertet}, \citenamefont {Ouerdane}, \citenamefont {Michot}, \citenamefont
  {Goupil},\ and\ \citenamefont {Lecoeur}}]{Apertet2013}%
  \BibitemOpen
  \bibfield  {author} {\bibinfo {author} {\bibfnamefont {Y.}~\bibnamefont
  {Apertet}}, \bibinfo {author} {\bibfnamefont {H.}~\bibnamefont {Ouerdane}},
  \bibinfo {author} {\bibfnamefont {A.}~\bibnamefont {Michot}}, \bibinfo
  {author} {\bibfnamefont {C.}~\bibnamefont {Goupil}}, \ and\ \bibinfo {author}
  {\bibfnamefont {P.}~\bibnamefont {Lecoeur}},\ }\href
  {http://stacks.iop.org/0295-5075/103/i=4/a=40001} {\bibfield  {journal}
  {\bibinfo  {journal} {EPL (Europhysics Letters)}\ }\textbf {\bibinfo {volume}
  {103}},\ \bibinfo {pages} {40001} (\bibinfo {year} {2013})}\BibitemShut
  {NoStop}%
\bibitem [{\citenamefont {Correa}\ \emph {et~al.}(2014)\citenamefont {Correa},
  \citenamefont {Palao}, \citenamefont {Adesso},\ and\ \citenamefont
  {Alonso}}]{Correa2014}%
  \BibitemOpen
  \bibfield  {author} {\bibinfo {author} {\bibfnamefont {L.~A.}\ \bibnamefont
  {Correa}}, \bibinfo {author} {\bibfnamefont {J.~P.}\ \bibnamefont {Palao}},
  \bibinfo {author} {\bibfnamefont {G.}~\bibnamefont {Adesso}}, \ and\ \bibinfo
  {author} {\bibfnamefont {D.}~\bibnamefont {Alonso}},\ }\href {\doibase
  10.1103/PhysRevE.90.062124} {\bibfield  {journal} {\bibinfo  {journal} {Phys.
  Rev. E}\ }\textbf {\bibinfo {volume} {90}},\ \bibinfo {pages} {062124}
  (\bibinfo {year} {2014})}\BibitemShut {NoStop}%
\bibitem [{\citenamefont {Holubec}\ and\ \citenamefont
  {Ryabov}(2015)}]{Holubec2016}%
  \BibitemOpen
  \bibfield  {author} {\bibinfo {author} {\bibfnamefont {V.}~\bibnamefont
  {Holubec}}\ and\ \bibinfo {author} {\bibfnamefont {A.}~\bibnamefont
  {Ryabov}},\ }\href {\doibase 10.1103/PhysRevE.92.052125} {\bibfield
  {journal} {\bibinfo  {journal} {Phys. Rev. E}\ }\textbf {\bibinfo {volume}
  {92}},\ \bibinfo {pages} {052125} (\bibinfo {year} {2015})}\BibitemShut
  {NoStop}%
\bibitem [{\citenamefont {Gonzalez-Ayala}\ \emph {et~al.}(2017)\citenamefont
  {Gonzalez-Ayala}, \citenamefont {Calvo~Hern\'andez},\ and\ \citenamefont
  {Roco}}]{Roco2017}%
  \BibitemOpen
  \bibfield  {author} {\bibinfo {author} {\bibfnamefont {J.}~\bibnamefont
  {Gonzalez-Ayala}}, \bibinfo {author} {\bibfnamefont {A.}~\bibnamefont
  {Calvo~Hern\'andez}}, \ and\ \bibinfo {author} {\bibfnamefont {J.~M.~M.}\
  \bibnamefont {Roco}},\ }\href {\doibase 10.1103/PhysRevE.95.022131}
  {\bibfield  {journal} {\bibinfo  {journal} {Phys. Rev. E}\ }\textbf {\bibinfo
  {volume} {95}},\ \bibinfo {pages} {022131} (\bibinfo {year}
  {2017})}\BibitemShut {NoStop}%
\bibitem [{\citenamefont {Johal}(2017{\natexlab{a}})}]{Johal2017}%
  \BibitemOpen
  \bibfield  {author} {\bibinfo {author} {\bibfnamefont {R.~S.}\ \bibnamefont
  {Johal}},\ }\href {\doibase 10.1103/PhysRevE.96.012151} {\bibfield  {journal}
  {\bibinfo  {journal} {Phys. Rev. E}\ }\textbf {\bibinfo {volume} {96}},\
  \bibinfo {pages} {012151} (\bibinfo {year} {2017}{\natexlab{a}})}\BibitemShut
  {NoStop}%
\bibitem [{\citenamefont {Angulo‐Brown}(1991)}]{Angulo1991}%
  \BibitemOpen
  \bibfield  {author} {\bibinfo {author} {\bibfnamefont {F.}~\bibnamefont
  {Angulo‐Brown}},\ }\href {\doibase 10.1063/1.347562} {\bibfield  {journal}
  {\bibinfo  {journal} {Journal of Applied Physics}\ }\textbf {\bibinfo
  {volume} {69}},\ \bibinfo {pages} {7465} (\bibinfo {year}
  {1991})}\BibitemShut {NoStop}%
\bibitem [{\citenamefont {Bejan}(1995)}]{Bejan1995}%
  \BibitemOpen
  \bibfield  {author} {\bibinfo {author} {\bibfnamefont {A.}~\bibnamefont
  {Bejan}},\ }\href@noop {} {\emph {\bibinfo {title} {Entropy Generation
  Minimization: The Method of Thermodynamic Optimization of Finite-Size Systems
  and Finite-Time Processes}}}\ (\bibinfo  {publisher} {CRC Press},\ \bibinfo
  {year} {1995})\BibitemShut {NoStop}%
\bibitem [{\citenamefont {Johal}(2017{\natexlab{b}})}]{Johalepjst}%
  \BibitemOpen
  \bibfield  {author} {\bibinfo {author} {\bibfnamefont {R.~S.}\ \bibnamefont
  {Johal}},\ }\href@noop {} {\bibfield  {journal} {\bibinfo  {journal} {The
  European Physical Journal Special Topics}\ }\textbf {\bibinfo {volume}
  {226}},\ \bibinfo {pages} {489} (\bibinfo {year}
  {2017}{\natexlab{b}})}\BibitemShut {NoStop}%
\bibitem [{\citenamefont {Reitlinger}(2008)}]{Reitlinger}%
  \BibitemOpen
  \bibfield  {author} {\bibinfo {author} {\bibfnamefont {H.~B.}\ \bibnamefont
  {Reitlinger}},\ }\href@noop {} {\emph {\bibinfo {title} {{\it Sur
  l'utilisation de la chaleur dans les machines \'{a} feu} ("On the use of heat
  in steam engines", in French).}}}\ (\bibinfo  {publisher}
  {Vaillant-Carmanne},\ \bibinfo {address} {Li\'{e}ge, Belgium},\ \bibinfo
  {year} {2008})\BibitemShut {NoStop}%
\bibitem [{\citenamefont {Chambadal}(1957)}]{Chambadal}%
  \BibitemOpen
  \bibfield  {author} {\bibinfo {author} {\bibfnamefont {P.}~\bibnamefont
  {Chambadal}},\ }\href@noop {} {\bibfield  {journal} {\bibinfo  {journal}
  {Armand Colin, Paris, France}\ }\textbf {\bibinfo {volume} {4}},\ \bibinfo
  {pages} {1} (\bibinfo {year} {1957})}\BibitemShut {NoStop}%
\bibitem [{\citenamefont {Novikov}(1958)}]{Novikov}%
  \BibitemOpen
  \bibfield  {author} {\bibinfo {author} {\bibfnamefont {I.}~\bibnamefont
  {Novikov}},\ }\href@noop {} {\bibfield  {journal} {\bibinfo  {journal} {J.
  Nucl. Energy II}\ }\textbf {\bibinfo {volume} {7}},\ \bibinfo {pages} {125}
  (\bibinfo {year} {1958})}\BibitemShut {NoStop}%
\bibitem [{\citenamefont {Tu}(2008)}]{Tu2008}%
  \BibitemOpen
  \bibfield  {author} {\bibinfo {author} {\bibfnamefont {Z.~C.}\ \bibnamefont
  {Tu}},\ }\href {http://stacks.iop.org/1751-8121/41/i=31/a=312003} {\bibfield
  {journal} {\bibinfo  {journal} {Journal of Physics A: Mathematical and
  Theoretical}\ }\textbf {\bibinfo {volume} {41}},\ \bibinfo {pages} {312003}
  (\bibinfo {year} {2008})}\BibitemShut {NoStop}%
\bibitem [{\citenamefont {Johal}\ and\ \citenamefont
  {Rai}(2016)}]{JohalRai2016}%
  \BibitemOpen
  \bibfield  {author} {\bibinfo {author} {\bibfnamefont {R.~S.}\ \bibnamefont
  {Johal}}\ and\ \bibinfo {author} {\bibfnamefont {R.}~\bibnamefont {Rai}},\
  }\href@noop {} {\bibfield  {journal} {\bibinfo  {journal} {EPL (Europhys.
  Lett.)}\ }\textbf {\bibinfo {volume} {113}},\ \bibinfo {pages} {10006}
  (\bibinfo {year} {2016})}\BibitemShut {NoStop}%
\bibitem [{\citenamefont {Bullen}(2003)}]{Bullen2003}%
  \BibitemOpen
  \bibfield  {author} {\bibinfo {author} {\bibfnamefont {P.~S.}\ \bibnamefont
  {Bullen}},\ }\href@noop {} {\emph {\bibinfo {title} {Handbook of Means and
  Their Inequalities}}}\ (\bibinfo  {publisher} {Kluwer},\ \bibinfo {address}
  {Dordrecht, Netherlands},\ \bibinfo {year} {2003})\BibitemShut {NoStop}%
  \bibitem [{\citenamefont {Esposito}\ \emph {et~al.}(2009)\citenamefont
  {Esposito}, \citenamefont {Lindenberg},\ and\ \citenamefont {Van~den
  Broeck}}]{Lindenberg2009}%
  \BibitemOpen
  \bibfield  {author} {\bibinfo {author} {\bibfnamefont {M.}~\bibnamefont
  {Esposito}}, \bibinfo {author} {\bibfnamefont {K.}~\bibnamefont
  {Lindenberg}}, \ and\ \bibinfo {author} {\bibfnamefont {C.}~\bibnamefont
  {Van~den Broeck}},\ }\href {\doibase 10.1103/PhysRevLett.102.130602}
  {\bibfield  {journal} {\bibinfo  {journal} {Phys. Rev. Lett.}\ }\textbf
  {\bibinfo {volume} {102}},\ \bibinfo {pages} {130602} (\bibinfo {year}
  {2009})}\BibitemShut {NoStop}%
  %
  \bibitem{Callenbook} H. B. Callen, {\it Thermodynamics and an Introduction
  to Thermostatistics}, Second Ed. (John Wiley and Sons \& Inc. U.K. 1985),
  Chapter 14.
 %
\bibitem [{\citenamefont {de~Groot}\ and\ \citenamefont
  {Mazur}(1969)}]{Grootbook}%
  \BibitemOpen
  \bibfield  {author} {\bibinfo {author} {\bibfnamefont {S.~R.}\ \bibnamefont
  {de~Groot}}\ and\ \bibinfo {author} {\bibfnamefont {P.}~\bibnamefont
  {Mazur}},\ }\href@noop {} {\emph {\bibinfo {title} {Non-equilibrium
  Thermodynamics}}}\ (\bibinfo  {publisher} {North-Holland Publishing
  Company},\ \bibinfo {address} {Amsterdam-London},\ \bibinfo {year}
  {1969})\BibitemShut {NoStop}%
%
\bibitem [{\citenamefont {Orlov}(1985)}]{Orlov}%
  \BibitemOpen
  \bibfield  {author} {\bibinfo {author} {\bibfnamefont {V.~N.}\ \bibnamefont
  {Orlov}},\ }\href@noop {} {\bibfield  {journal} {\bibinfo  {journal} {Sov.
  Phys. Dokl.}\ }\textbf {\bibinfo {volume} {30}},\ \bibinfo {pages} {506}
  (\bibinfo {year} {1985})}\BibitemShut {NoStop}%
\bibitem [{\citenamefont {Hardy}\ \emph {et~al.}(1952)\citenamefont {Hardy},
  \citenamefont {Littlewood},\ and\ \citenamefont {P\'{o}lya}}]{Hardy1952}%
  \BibitemOpen
  \bibfield  {author} {\bibinfo {author} {\bibfnamefont {G.~H.}\ \bibnamefont
  {Hardy}}, \bibinfo {author} {\bibfnamefont {J.~E.}\ \bibnamefont
  {Littlewood}}, \ and\ \bibinfo {author} {\bibfnamefont {G.}~\bibnamefont
  {P\'{o}lya}},\ }\href@noop {} {\emph {\bibinfo {title} {Inequalities}}}\
  (\bibinfo  {publisher} {Cambridge University Press},\ \bibinfo {address}
  {Cambridge},\ \bibinfo {year} {1952})\BibitemShut {NoStop}%
\bibitem [{\citenamefont {Cavina}\ \emph {et~al.}(2017)\citenamefont {Cavina},
  \citenamefont {Mari},\ and\ \citenamefont {Giovannetti}}]{Cavina2017}%
  \BibitemOpen
  \bibfield  {author} {\bibinfo {author} {\bibfnamefont {V.}~\bibnamefont
  {Cavina}}, \bibinfo {author} {\bibfnamefont {A.}~\bibnamefont {Mari}}, \ and\
  \bibinfo {author} {\bibfnamefont {V.}~\bibnamefont {Giovannetti}},\ }\href
  {\doibase 10.1103/PhysRevLett.119.050601} {\bibfield  {journal} {\bibinfo
  {journal} {Phys. Rev. Lett.}\ }\textbf {\bibinfo {volume} {119}},\ \bibinfo
  {pages} {050601} (\bibinfo {year} {2017})}\BibitemShut {NoStop}%
\bibitem [{\citenamefont {Yan}(1993)}]{Yan1993comment}%
  \BibitemOpen
  \bibfield  {author} {\bibinfo {author} {\bibfnamefont {Z.}~\bibnamefont
  {Yan}},\ }\href {\doibase 10.1063/1.354041} {\bibfield  {journal} {\bibinfo
  {journal} {Journal of Applied Physics}\ }\textbf {\bibinfo {volume} {73}},\
  \bibinfo {pages} {3583} (\bibinfo {year} {1993})},
  \BibitemShut {NoStop}%
\bibitem [{\citenamefont {Hern\'andez}\ \emph {et~al.}(2001)\citenamefont
  {Hern\'andez}, \citenamefont {Medina}, \citenamefont {Roco}, \citenamefont
  {White},\ and\ \citenamefont {Velasco}}]{Calvo2001}%
  \BibitemOpen
  \bibfield  {author} {\bibinfo {author} {\bibfnamefont {A.~C.}\ \bibnamefont
  {Hern\'andez}}, \bibinfo {author} {\bibfnamefont {A.}~\bibnamefont {Medina}},
  \bibinfo {author} {\bibfnamefont {J.~M.~M.}\ \bibnamefont {Roco}}, \bibinfo
  {author} {\bibfnamefont {J.~A.}\ \bibnamefont {White}}, \ and\ \bibinfo
  {author} {\bibfnamefont {S.}~\bibnamefont {Velasco}},\ }\href {\doibase
  10.1103/PhysRevE.63.037102} {\bibfield  {journal} {\bibinfo  {journal} {Phys.
  Rev. E}\ }\textbf {\bibinfo {volume} {63}},\ \bibinfo {pages} {037102}
  (\bibinfo {year} {2001})}\BibitemShut {NoStop}%
\bibitem [{\citenamefont {Long}\ and\ \citenamefont {Liu}(2015)}]{Long2015}%
  \BibitemOpen
  \bibfield  {author} {\bibinfo {author} {\bibfnamefont {R.}~\bibnamefont
  {Long}}\ and\ \bibinfo {author} {\bibfnamefont {W.}~\bibnamefont {Liu}},\
  }\href@noop {} {\bibfield  {journal} {\bibinfo  {journal} {Phys. Lett. A}\
  }\textbf {\bibinfo {volume} {434}},\ \bibinfo {pages} {232} (\bibinfo {year}
  {2015})}\BibitemShut {NoStop}%
\bibitem [{\citenamefont {Yan}\ and\ \citenamefont {Chen}(1990)}]{Chen1990}%
  \BibitemOpen
  \bibfield  {author} {\bibinfo {author} {\bibfnamefont {Z.}~\bibnamefont
  {Yan}}\ and\ \bibinfo {author} {\bibfnamefont {J.}~\bibnamefont {Chen}},\
  }\href {http://stacks.iop.org/0022-3727/23/i=2/a=002} {\bibfield  {journal}
  {\bibinfo  {journal} {Journal of Physics D: Applied Physics}\ }\textbf
  {\bibinfo {volume} {23}},\ \bibinfo {pages} {136} (\bibinfo {year}
  {1990})}\BibitemShut {NoStop}%
\bibitem [{\citenamefont {de~Tomas}\ \emph {et~al.}(2013)\citenamefont
  {de~Tomas}, \citenamefont {Roco}, \citenamefont {Hern\'andez}, \citenamefont
  {Wang},\ and\ \citenamefont {Tu}}]{Calvo2013}%
  \BibitemOpen
  \bibfield  {author} {\bibinfo {author} {\bibfnamefont {C.}~\bibnamefont
  {de~Tomas}}, \bibinfo {author} {\bibfnamefont {J.~M.~M.}\ \bibnamefont
  {Roco}}, \bibinfo {author} {\bibfnamefont {A.~C.}\ \bibnamefont
  {Hern\'andez}}, \bibinfo {author} {\bibfnamefont {Y.}~\bibnamefont {Wang}}, \
  and\ \bibinfo {author} {\bibfnamefont {Z.~C.}\ \bibnamefont {Tu}},\ }\href
  {\doibase 10.1103/PhysRevE.87.012105} {\bibfield  {journal} {\bibinfo
  {journal} {Phys. Rev. E}\ }\textbf {\bibinfo {volume} {87}},\ \bibinfo
  {pages} {012105} (\bibinfo {year} {2013})}\BibitemShut {NoStop}%
\bibitem [{\citenamefont {Johal}(2016)}]{Johal2016}%
  \BibitemOpen
  \bibfield  {author} {\bibinfo {author} {\bibfnamefont {R.~S.}\ \bibnamefont
  {Johal}},\ }\href@noop {} {\bibfield  {journal} {\bibinfo  {journal} {Phys.
  Rev. E}\ }\textbf {\bibinfo {volume} {94}},\ \bibinfo {pages} {012123}
  (\bibinfo {year} {2016})}\BibitemShut {NoStop}%

\end{thebibliography}
\end{document}